\documentclass[twocolumn,aps,prl,floats,showpacs]{revtex4}
\usepackage{epsfig}
\usepackage{dcolumn}
\usepackage{bm}
\begin{document}

\title{Nonequilibrium Relaxations and Aging Effects in a Two-Dimensional Coulomb Glass}
\author{J. Jaroszy\'nski}
\affiliation{National High Magnetic Field Laboratory, Florida State
University, Tallahassee, Florida 32310}
\author{Dragana  Popovi\'c}
\email{dragana@magnet.fsu.edu} \affiliation{National High Magnetic Field Laboratory, Florida State University,
Tallahassee, Florida 32310}
\date{\today}

\begin{abstract}
The relaxations of conductivity have been studied in the glassy regime of a strongly disordered two-dimensional
electron system in Si after a temporary change of carrier density during the waiting time $t_w$. Two types of
response have been observed: a) monotonic, where relaxations exhibit aging, \textit{i.e.} dependence on history,
determined by $t_w$ and temperature; b) nonmonotonic, where a memory of the sample history is lost.  The
conditions that separate the two regimes have been also determined.
\end{abstract}


\maketitle

The dynamics of glasses and other systems out of equilibrium
is one of the most challenging and rapidly evolving topics in condensed matter research \cite{LesHouches-2002}.
Although glassy behavior may dominate the low-temperature properties of many complex materials near quantum
phase transitions~\cite{Mir-Dob-review}, such as the metal-insulator transition (MIT), quantum glasses are
less understood than their classical counterparts.  Experimental studies of
charge or Coulomb glasses~\cite{eglass}, which are of particular relevance to the MIT, have been
scarce~\cite{earlyg,films-Zvi,films3}, and limited to insulating systems far from the MIT. Recent
observations~\cite{SBPRL,JJPRL,relax-PRL} of
glassiness in a two-dimensional electron system (2DES) in Si open up opportunities for exploring glassy
phenomena over a wide range of all the relevant parameters.  The carrier density $n_s$, for example, can be
varied continuously on the same sample from the deep metallic into the deep insulating regime. The onset of
glassiness was, in fact, found~\cite{SBPRL,JJPRL,relax-PRL} to take place at a density $n_g>n_c$ ($n_c$ -- the
critical density for the MIT), \textit{i.e.} on the metallic side of the MIT, consistent with
theory~\cite{Darko}.

The manifestations of glassiness in a 2DES for $n_s<n_g$, as demonstrated by resistance noise
measurements~\cite{SBPRL,JJPRL}, include a dramatic slowing down of the electron dynamics and correlated
statistics consistent with the hierarchical picture of glasses.  Furthermore,
studies of the relaxations of the conductivity $\sigma$~\cite{relax-PRL}, following a rapid
change of $n_s$,
show that the equilibration time $\tau_{eq}$~\cite{tauhigh} diverges exponentially as temperature $T\rightarrow
0$, suggesting a glass transition at $T_{g}=0$.  However, even at $T$ that are not too low (\textit{e.g.} $\sim
1$~K), $\tau_{eq}$ exceeds easily not only the experimental time window but also the age of the Universe.  This
makes it relatively easy to study the out-of-equilibrium relaxation of $\sigma$ at times $t\ll\tau_{eq}$, where
one expects to find properties common to other types of glasses. Indeed, for $t\ll\tau_{eq}$, the relaxations
obey~\cite{relax-PRL} $\sigma(t,T,n_s)/\sigma_{0}(T,n_s)\propto
t^{-\alpha(n_s)}\exp[-(t/\tau(T,n_s))^{\beta(n_s)}]$ with $\tau$ diverging as $T\rightarrow 0$
($\sigma_{0}(T,n_s)$ -- equilibrium $\sigma$ at a given $T$ and $n_s$; $0<\alpha(n_s)<0.4$,
$0.2<\beta(n_s)<0.45$). Such scaling has been observed in spin glasses above $T_g$~\cite{Pappas}. The
dependence $\tau\propto\exp(\gamma n_{s}^{1/2})$ ($\gamma$ -- a proportionality constant)~\cite{relax-PRL} is a
strong indication for the relevance of Coulomb interactions between 2D electrons~\cite{Coulomb-comment} in the
$n_s<n_g$ regime.  The most puzzling feature of the relaxation in
Ref.~\cite{relax-PRL}, however, is that the system equilibrates only after it first goes farther away from
equilibrium. This ``overshooting'' (OS) of $\sigma_0$ is discussed in more detail below.

A key characteristic of relaxing glassy systems is the loss of time translation invariance, reflected in
\textit{aging} effects~\cite{Struik,aging}.  The system is said to exhibit aging if its response to an external
excitation depends on the system history in addition to the time $t$.  Some history dependence was observed in a
2DES~\cite{SBPRL}, but there have been no systematic and detailed studies. Here we report the first such study,
where, for a given $n_s$, $\sigma(t)$ was measured after the temporary change of $n_s$ during the waiting time
$t_w$.  The measurement history was varied systematically by changing $t_w$ and $T$.  Two types of response have
been observed: a) \textit{monotonic}, for relatively ``small'' excitations, where relaxations depend on $t_w$,
\textit{i.e.} the 2DES exhibits aging; b) \textit{nonmonotonic}, for sufficiently ``large'' excitations, where
$\sigma(t)$ overshoots $\sigma_0$ and relaxations no longer depend on $t_w$ (memory loss).  The conditions that
separate the two regimes have been identified and discussed.

Measurements were carried out on a 2DES in Si metal-oxide-semiconductor field-effect transistors
with a 50~nm oxide thickness, and a relatively large amount of disorder (the 4.2~K peak mobility
$\approx$~0.06~m$^2$/Vs with the substrate (back-gate) bias~\cite{AFS} $V_{sub}=-2$~V). The sample
length$\,\times\,$width were $1\times 90~\mu$m$^2$ (sample A) and $2\times 50~\mu$m$^2$ (sample B). The same
devices were used in Ref.~\cite{relax-PRL}, and they were also very similar to the samples used in noise
studies~\cite{SBPRL}. The density $n_s$ is controlled by the gate voltage $V_g$, such that
$n_s(10^{11}$cm$^{-2})=4.31(V_g[$V$]-V_{T})$, where the threshold voltage for forming the 2D layer $V_T=6.3$~V
for $V_{sub}=-2$~V; $n_g(10^{11}$cm$^{-2})\approx 7.5$ and $n_c(10^{11}$cm$^{-2})\approx 4.5$~\cite{relax-PRL}.
The samples and the standard ac lock-in technique (typically 13 Hz) that was used to measure $\sigma$ have been
described in more detail elsewhere~\cite{SBPRL}.

\begin{figure}
\centerline{\epsfig{file=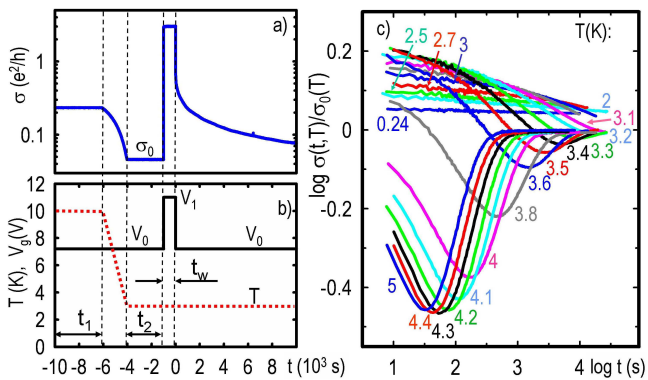,width=8.5cm,clip=}}\caption{(Color online) (a) Sample A.  $\sigma(t)$ for
$V_{0}=7.2$~V [$n_s(10^{11}$cm$^{-2})=3.88<n_c$], $V_{1}=11$~V [$n_s(10^{11}$cm$^{-2})=20.26$], $t_{w}=1000$~s,
and $T=3.0$~K. (b) The corresponding experimental protocol: $V_g(t)$ and $T(t)$. The results do not depend on
the cooling time (varied from 30 minutes to 10 hours), nor on
$t_1$ and $t_2$ (varied from 5 minutes to 8 hours each).  (c) Sample B.  $\sigma(t)$ for $V_{0}=8$~V
[$n_s(10^{11}$cm$^{-2})=7.33\lesssim n_g$], $V_{1}=11$~V, $t_{w}=1000$~s, and several $T$, as shown.  $T=2.3,
1.8, 1.3, 1.0, 0.75, 0.5$~K data
were omitted for clarity.\label{fig:example2}}
\end{figure}

The experimental protocol [Figs.~\ref{fig:example2}(a) and \ref{fig:example2}(b)] starts with the 2DES, induced
by an applied gate voltage
%
%
$V_{0}$, in equilibrium at 10~K.  The sample is then cooled to the measurement temperature $T$.  Unlike glasses
with $T_{g}\neq 0$, this does not result in any visible relaxations on our experimental time scales, and
$\sigma_0(V_{0},T)$ is thus established. $V_g$ is then switched rapidly (within 1~s) to a different value
$V_{1}$, where it is kept for a time $t_w$. Finally, it is changed back
(within 1~s) to $V_0$, and the
slowly evolving $\sigma(t,V_{0},T)$ is measured. We note that $t=0$ is defined as the time when $V_g$ reattains
its original value $V_0$.  An analogous protocol has been used often in studies of other glassy
materials~\cite{films-Zvi}.  In general, we expect the larger values
of $t_w$ or $T$ to correspond to stronger excitations of the 2DES, because the system will have more time or
more thermal energy, respectively, to wander away from its original equilibrium state during the time $t_w$.

Figure~\ref{fig:example2}(c) shows some relaxations measured for a fixed $t_w$ at different $T$ for a given
$V_0$ and $V_1$.  The data are not shown for the first few seconds (comparable to our sampling time $\sim 1$~s),
when $\sigma$ changes rapidly, consistent with the $\sigma_{0}(V_g)$ dependence~\cite{AFS}.  In general, the
observed behavior is very complicated, but it can be divided roughly into two types. At higher $T$, $\sigma(t)$
first goes below $\sigma_0$, and then even farther away from equilibrium. Eventually, it changes ``direction''
and approaches $\sigma_0$.  Similar nonmonotonic behavior was observed after exciting 2DES out of equilibrium by
a rapid change of $n_s$~\cite{relax-PRL}.  At lower $T$, on the other hand, there is no
OS of equilibrium and $\sigma$ approaches $\sigma_{0}$ monotonically, similar to observations in other electron
glasses~\cite{films-Zvi}.

\begin{figure*}
\centerline{\epsfig{file=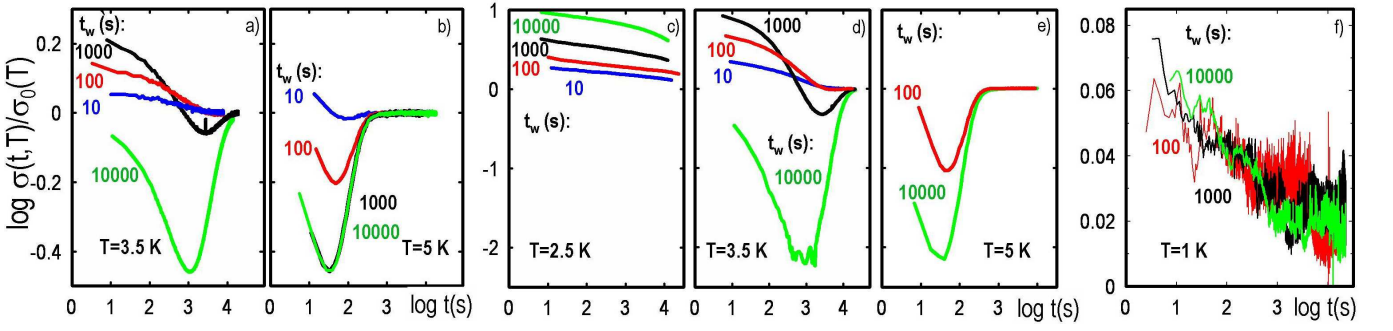,width=18.0cm,clip=}} \caption{(Color online) $\sigma(t)$ for different
$t_{w}$ at fixed $T$, as shown.  (a)-(b) Sample B; $V_{0}=8$~V, $V_{1}=11$~V; (c)-(e) Sample B; $V_{0}=7.2$~V,
$V_{1}=11$~V.  (f) Sample A; $V_{0}=7.2$~V and $V_{1}=6$~V$(n_s=0)<V_{T}=6.3$~V, \textit{i.e.} all electrons are
removed from the 2D layer during $t_w$.
 \label{fig:diffT}}
\end{figure*}

$\sigma(t)$ were also measured for different $t_w$ at a fixed $T$ and given $V_0$ and $V_1$.  Some typical data,
including those for two different $V_0$, are shown in Fig.~\ref{fig:diffT}. In general, the
%
%
waiting time $t_w$ has a dramatic effect on the relaxations, both qualitatively (OS \textit{vs.} no OS) and
quantitatively (\textit{e.g.} more than two orders of magnitude difference in $\sigma(t)$ in
Fig.~\ref{fig:diffT}(d)), including instances of OS when $t_w$-dependence vanishes [Fig.~\ref{fig:diffT}(b)].
$\sigma(t)$ were also investigated
after \textit{all} the electrons have been removed from the 2D layer during the time $t_w$.  In a 2DES in Si,
unlike other electron glasses, this may be accomplished
simply by reducing $V_g$ by a sufficient amount.  If the slow dynamics is dominated by intrinsically glassy
behavior of the interacting electrons rather than by their response to extrinsic slow degrees of freedom
(\textit{e.g.} motion of nearby impurities, although exponentially unlikely at our $T$~\cite{AFS}), one expects
that $\sigma(t)$, measured after the electrons are reintroduced with the initial $V_g=V_0$, will not depend on
$t_w$. The measurements confirm this expectation [Fig.~\ref{fig:diffT}(f)], providing further support for the
glassiness in the 2DES.

\begin{figure}
\centerline{\epsfig{file=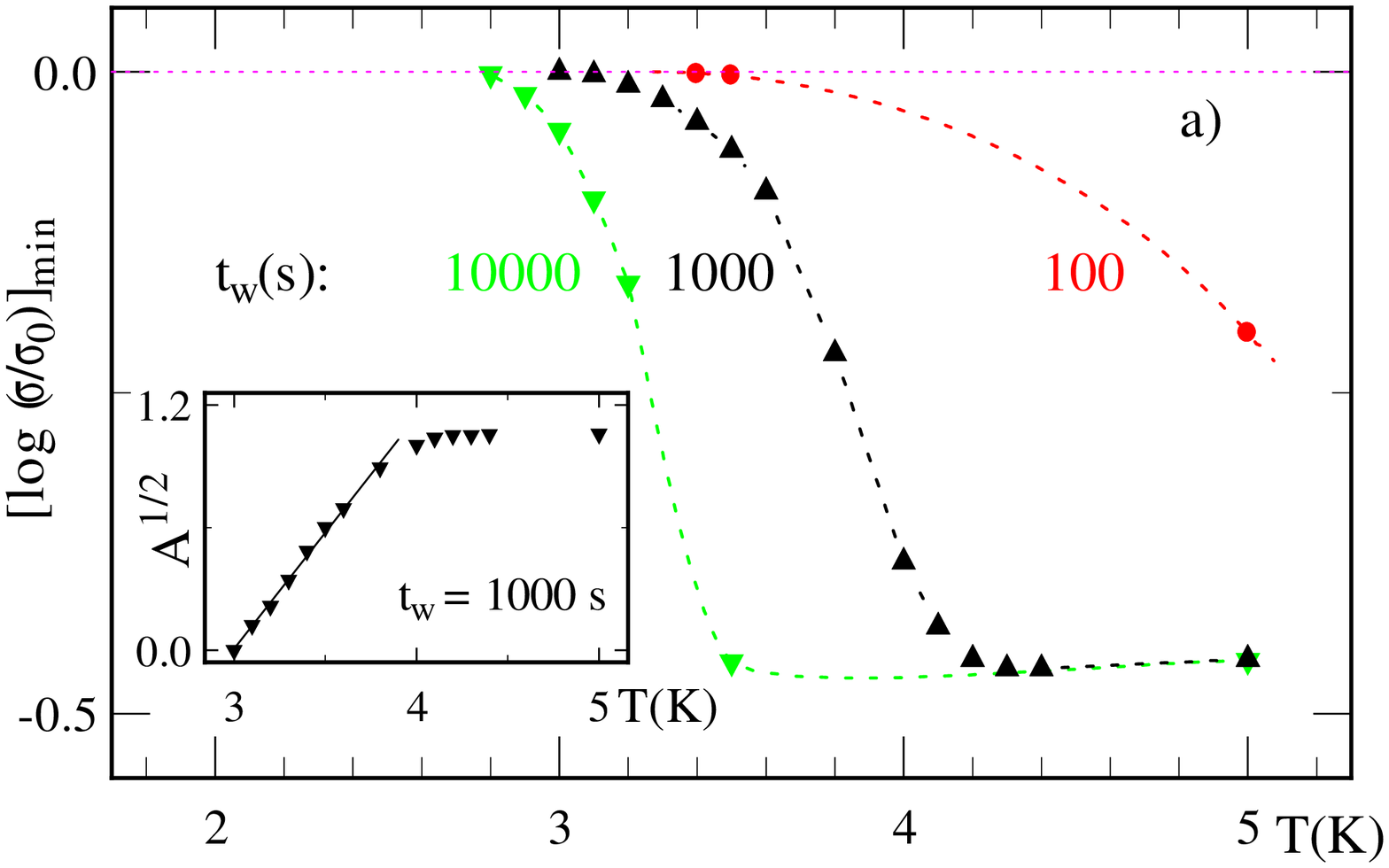,width=8.5cm,clip=}} \vspace*{-10pt}
\centerline{\epsfig{file=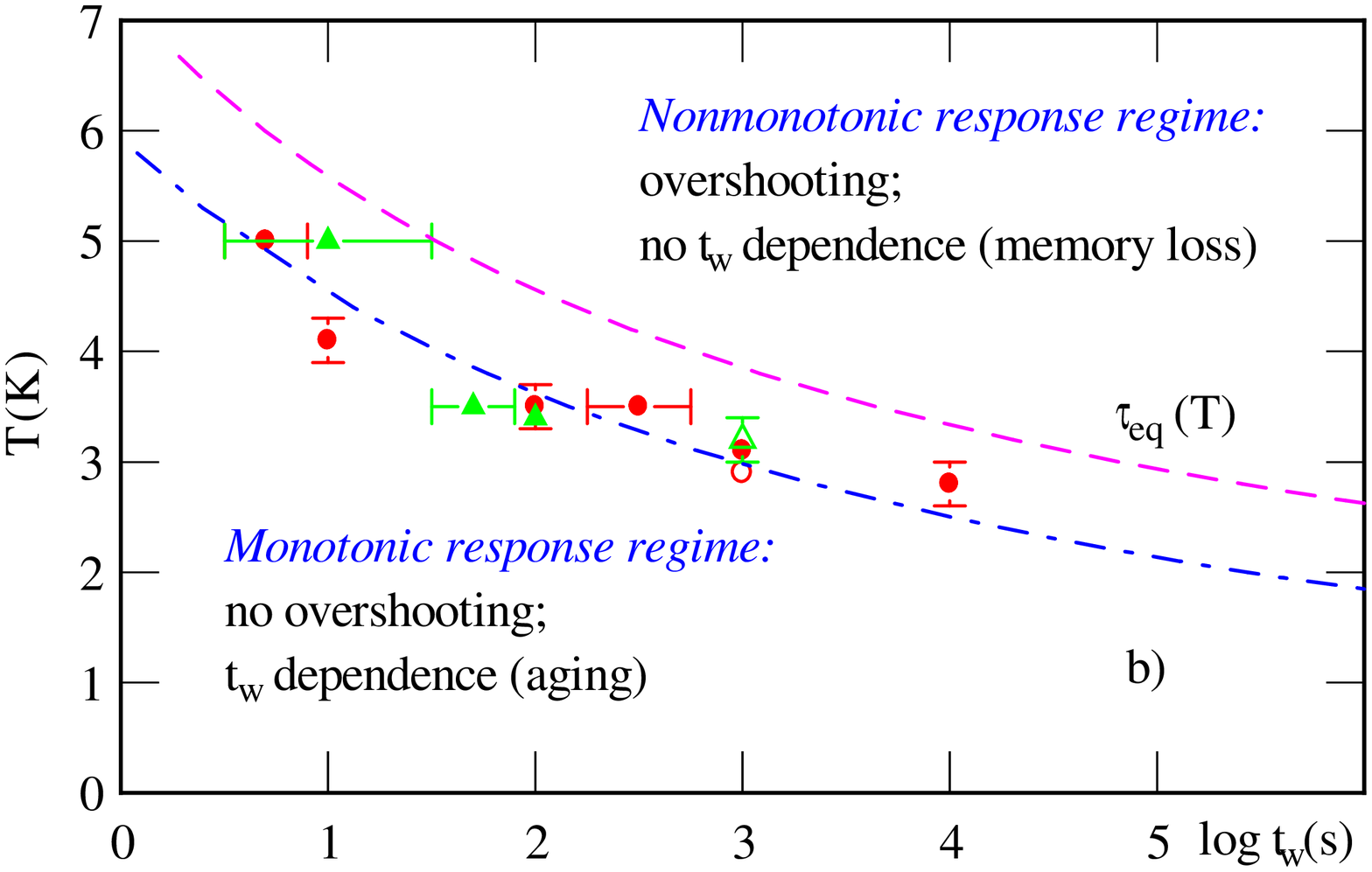,width=8.5cm,clip=}} \caption{(Color online) (a)
$\log(\sigma/\sigma_0)$ at the $\sigma(t)$ minima (data with overshooting) \textit{vs.} $T$ for different $t_w$.
Sample B; $V_0=8$~V, $V_1=11$~V. The lines guide the eye.  Inset: $A(T)$ for $t_w=1000$~s.  The solid line is a
linear fit.  (b) The symbols (solid and open from devices B and A, respectively) show the values of $(T,t_w)$
where the overshooting vanishes. Red dots and green triangles correspond to $V_0=8$~V, $V_1=11$~V, and
$V_0=7.2$~V, $V_1=11$~V, respectively. The blue dot-dashed line guides the eye.  The purple dashed line
represents $\tau_{eq}(T)$.\label{fig:phasediagram}}
\end{figure}

A detailed analysis of $\sigma(t)$ in the case of OS [\textit{e.g.} highest $T$ curves in
Fig.~\ref{fig:example2}(c)] finds the behavior very similar to that reported in Ref.~\cite{relax-PRL}.
Therefore, we do not show the fits to the data, but instead summarize the main results.  For example, at
$t$ just before the minimum in $\sigma(t)$~\cite{comment-max},
$\sigma(t,T,n_s)/\sigma_0(T,n_s)\propto\exp[-(t/\tau(T,n_s))^{\beta(n_s)}]$, with $\tau$ and
$\beta$~\cite{comment-exponents} similar to those in Ref.~\cite{relax-PRL}. After the minimum in $\sigma(t)$,
the approach to equilibrium at the longest
$t$ is described by a simple exponential process $\sigma/\sigma_0=[1-A(T)\exp(-t/\tau_{eq})]$ with
$1/\tau_{eq}\propto\exp(-E_A/T)$ ($E_A\approx57$~K as
before~\cite{relax-PRL,tauhigh}).  The main difference between the equilibrations that follow the two
experimental protocols is that the prefactor $A$ did not depend on $T$ in Ref.~\cite{relax-PRL}, whereas here
$A(T)$ decreases with decreasing $T$ and vanishes at a finite $T$ (Fig.~\ref{fig:phasediagram}(a) inset).
%
%
This, at least partly, reflects the observation that, in Ref.~\cite{relax-PRL}, the minima in the $\sigma(t)$
curves only move to lower values of $\sigma$ and to later $t$ with decreasing $T$, whereas in the current
protocol, the different $T$ curves at $t$ after the minimum [Fig.~\ref{fig:example2}(c)] are no longer parallel
to each other, the minima become ``shallower'', and eventually vanish at a finite $T$.  The latter is shown in
Fig.~\ref{fig:phasediagram}(a), where the $T$-dependences of the amplitude of the relaxations at the minima,
defined as $[\log(\sigma/\sigma_0)]_{min}$, are plotted for different $t_w$.  The amplitudes extrapolate to zero
at $T$ that are in a good agreement with those obtained from $A(T)$.  Thus the data seem to indicate that here
the OS does not simply fall out of our experimental time window at low $T$ as in Ref.~\cite{relax-PRL} but
rather that it vanishes at a finite $T$. The $\sigma(t)$ curves at lower $T$ support this conclusion: their
monotonic approach to $\sigma_0$ [from above in Fig.~\ref{fig:example2}(c)] is consistent with a power-law form
at the shortest $t$ (lowest $T$) and, at the longest $t$, with a simple exponential process with the same
$E_A\approx 57$~K.

The procedure shown in Fig.~\ref{fig:phasediagram}(a) was applied to all the data in order to determine the
values of $(T,t_w)$ where OS disappears.  The results [Fig.~\ref{fig:phasediagram}(b)] clearly divide the phase
space in two, such that the OS is not observed for smaller values of $(T,t_w)$ (\textit{e.g.} at 1~K, not even
at $t\simeq 2\times 10^5$~s).  This confirms our earlier speculation~\cite{relax-PRL} that the OS occurs for
sufficiently strong perturbations out of equilibrium.  We note that
the relevant time (energy)
scale is given by $\tau_{eq}(T)$ [Fig.~\ref{fig:phasediagram}(b)].  In order to understand
this result better, we discuss the experiment
in more detail.  During
$t_w$, after $V_g$ is
switched rapidly from $V_0$ to $V_1$, $\sigma(t,V_1,T)$ evolves with
$t$ as described in Ref.~\cite{relax-PRL}. Some small amplitude relaxations are still visible [albeit not on the
scale of Fig.~\ref{fig:example2}(a)] even at the highest $V_1=11$~V where
$n_s>n_g$, but $\sigma_0$ is still so small that $k_{F}l<1$ ($k_{F}$ -- Fermi wave vector, $l$ -- mean free
path)~\cite{kfl}. If the 2DES equilibrates at a new state corresponding to $V_1$, \textit{i.e.} if $\tau_{eq}\ll
t_w$
($\tau_{eq}$ does not depend on $n_s$ \cite{relax-PRL}), then the rest of
this protocol, namely the switching of $V_1$ back to $V_0$ and the subsequent relaxation, are equivalent to the
protocol employed in Ref.~\cite{relax-PRL}.  Hence
both here and in Ref.~\cite{relax-PRL}, \textit{the OS is
observed only when the 2DES is excited out of thermal equilibrium}. Of course, if $\tau_{eq}\ll t_w$, then any
further increase in $t_w$ will have no effect on
$\sigma(t)$, as indeed observed [Fig.~\ref{fig:diffT}(b)]. On the other hand, when $t_w\ll\tau_{eq}$ [monotonic
response regime in Fig.~\ref{fig:phasediagram}(b)],
$\sigma(t)$ clearly depend on history (aging effect), \textit{i.e.} on
$t_w$ [see \textit{e.g.}
Fig.~\ref{fig:diffT}(c)] during which the system relaxes away from its initial equilibrium state determined by
$V_0$ and towards a new equilibrium state determined by $V_1$. It can be also said that the system has a
\textit{memory} of the time it spent with $V_g=V_1$.
This is very similar to
spin glasses, for example, where $T$ or a magnetic field play a role of $n_s$
in our experiment. Finally, in the crossover regime $t_w\lesssim\tau_{eq}$ [a region between the two dashed
lines in Fig.~\ref{fig:phasediagram}(b)], both some OS and $t_w$-dependence are observed
[Fig.~\ref{fig:diffT}(d)].

The nonmonotonic response of $\sigma$, which occurs when the system is taken out of thermal equilibrium, has
been observed for all applied $1\leq\mid\!\!\Delta V_g\!\!\mid =\mid\!\!V_1-V_0\!\!\mid$~(V)$\leq 4.7$
($\mid\!\Delta n_s\!\mid/n_s\approx 20-100$\%).  Since, by reducing $n_s$, the 2DES goes from a metallic regime,
with no metastable states, to a glassy, and then glassy insulating regime, it seems reasonable to assume that
the density of \textit{metastable} states increases as $n_s$ is reduced by a large amount, reflecting the
increasing complexity of the free energy landscape.  The 2DES are also known to be relatively poorly coupled
thermally to the surrounding lattice, especially in Si, and thermal equilibrium at very low T (below
$\sim\!\!1$~K) is achieved mainly by conduction through the contacts and the leads. This and the available data
suggest that the rapid change of $V_g$
is an approximately adiabatic process.  In that
case, the reduction of $n_s$, accompanied by an increase in the density of metastable states, would lead to the
cooling of the 2DES. As a result, $\sigma$ would decrease with $t$ ($d\sigma/dT>0$ in the regime of interest)
below $\sigma_0(T)$. This initial cooling of the 2DES and the eventual warm-up to the bath $T$ would give rise
to a minimum in $\sigma(t)$ as observed. Likewise, a nearly adiabatic increase of $n_s$ would lead to the
heating of the 2DES,
resulting in a maximum in $\sigma(t)$~\cite{comment-max}.
The time-scale for equilibration would be determined both by heat leaks from the bath and by slow
processes~\cite{Pobell} related to collective rearrangements of the 2DES to adjust to the new conditions. The
applied $\mid\!\!\Delta V_g\!\!\mid$ are expected to trigger major rearrangements of the electron configuration
since the corresponding shifts of the Fermi energy $\Delta E_F\gg T$~\cite{Muller-French,Coulomb-comment}. This
situation, coupled with possibly substantial changes in the screening of the 2DES across the
MIT~\cite{screening}, is fundamentally different from the cooling of the 2DES at a fixed $n_s$ when $\Delta T\ll
E_F$.  Charge rearrangements have been also invoked to interpret the nonmonotonic relaxation in insulating
granular metals~\cite{Aviad-opposite}. The
cooling (heating) scenario proposed here is consistent with
the data but
other interpretations may be possible, especially since OS
is found in many systems, ranging from
manganites~\cite{mang-Levy} to biological systems~\cite{bio-Nelson}. For
example, OS
may be a general feature of systems that are
far from equilibrium~\cite{roundabout}.

The adiabatic cooling (heating) effects should not be relevant for $t_w\ll\tau_{eq}$, \textit{i.e.} for
``small'' excitations, since the final state will have a large configurational similarity with the original
state due to the shortness of $t_w$.  Such small perturbations of a Coulomb glass are hence expected to lead to
memory effects~\cite{Muller-memory}, all of which are consistent with our results.  Studies of other electron
glasses, where no OS but aging and memory effects were seen~\cite{films-Zvi}, were done in such a perturbative
regime.  An equivalent (deep in the $t_w\ll\tau_{eq}$ limit) study in the 2DES would be of great interest,
especially across the MIT, but that is beyond the scope of this paper.

Finally, we have attempted to determine the microscopic origin of the activation energy $E_A\approx 57$~K by
repeating some of the measurements for different $V_{sub}$. In addition to changing the disorder, $V_{sub}$
moves the position of the 2D subband with respect to the bottom of the Si conduction band, and it also affects
the splitting between the subbands of the 2DES~\cite{AFS,Feng-moments}.  If the exponential process at long $t$
results from an activation to Si-SiO$_2$ interface traps or to an upper subband, it can be
estimated~\cite{AFS,SBPRL} that the applied range $-5\leq V_{sub}$(V)$\leq 0$ will have a significant impact on
the value of $E_A$.  We find, however, no change in $E_A$, which seems to rule out the above two processes as
mechanisms for equilibration.

We have demonstrated that the 2DES in Si exhibits aging effects, one of the hallmarks of glassy behavior,
and identified precisely the conditions that lead to memory loss and nonmonotonic response. The observed complex
dynamics of the electronic transport is strikingly similar to that of
other
systems that are far from equilibrium.

We are grateful to I. Rai\v{c}evi\'{c} for technical assistance,
V. Dobrosavljevi\'c for useful discussions, NSF DMR-0403491 and
NHMFL via NSF Cooperative Agreement DMR-0084173 for financial
support. D.~P. also acknowledges the hospitality of the Aspen
Center for Physics.

\newcommand{\noopsort}[1]{} \newcommand{\printfirst}[2]{#1}
  \newcommand{\singleletter}[1]{#1} \newcommand{\switchargs}[2]{#2#1}


\begin{thebibliography}{24}
\bibitem[{tau()}]{tauhigh}
\bibinfo{note}{In Ref.~\cite{relax-PRL}, this time was denoted by
$\tau_{high}/k_{1}=(3.7\times 10^{-4}$~s$)\exp(E_A/T)$ ($E_A\approx57$~K independent of $n_s$)}.
\bibitem[{com()}]{comment-exponents}
\bibinfo{note}{$\tau$ and $\beta$ depend only on the final $n_s$, not on $(V_1-V_0)$}.
\bibitem[{kfl()}]{kfl}
\bibinfo{note}{Drude $\sigma=g_{v}(e^2/h)(k_{F}l)$; $g_v=2$ is a valley degeneracy.  None of the
effects described here are seen for $k_{F}l>1$}.
\end{thebibliography}
\end{document}